\documentclass[aps,pra,preprint,groupedaddress]{revtex4-1}
\usepackage{graphicx}%
\usepackage{dcolumn}%
\usepackage{bm}%

\begin{document}
\title{An optical equilibrium in lateral waveguide-resonator optical force}
\author{Varat Intaraprasonk}
\affiliation{Department of Applied Physics, Stanford University, Stanford, California 94305}

\author{Shanhui Fan}
\affiliation{Department of Electrical Engineering, Stanford University, Stanford, California 94305}

\date{\today}
\begin{abstract}
We consider the lateral optical force between resonator and waveguide, and study the possibility of an equilibrium that occurs solely from the optical force. We prove analytically that a single-resonance system cannot give such an equilibrium in resonator-waveguide force. We then show that two-resonance systems can provide such an equilibrium. We provide an intuitive way to predict the existence of an equilibrium, and give numerical examples.
\end{abstract}
\maketitle

\section{Introduction}

There have been significant recent interests on the optical force in photonic nano-structures \cite{vanthourhout,rakich2010, liu2009, pernice2009, roels2009, pernice2010, povinelli2005, li2009, varatmemory, einat2011, varateit, eichenfield2007, povinelli2005resres, rakich2007, rosenberg2009, wiederhecker2009, wiederhecker2011, li2009prl}, such as coupled waveguides system \cite{pernice2009, roels2009, pernice2010, povinelli2005, li2009, varatmemory}, coupled resonator-waveguide systems \cite{einat2011, varateit, eichenfield2007, li2009prl}, or coupled resonators systems \cite{povinelli2005resres, rakich2007, rosenberg2009, wiederhecker2009, wiederhecker2011}. In many cases, the optical force is strong enough to change the static mechanical configurations of photonic nanostructures \cite{povinelli2005, li2009, varatmemory, eichenfield2007, rakich2007, rosenberg2009, wiederhecker2009, wiederhecker2011}. As one set of examples we note the demonstrations of the bending of optical waveguides, either by the optical force between two parallel waveguides \cite{povinelli2005, li2009, varatmemory}, or the force between a resonator and a feeding waveguide \cite{eichenfield2007}. Another example is the change in the separation between two coupled resonators due to the optical force between them \cite{rakich2007, rosenberg2009, wiederhecker2009, wiederhecker2011}. These changes in mechanical configurations can be used to optically tune the systems' frequency response \cite{li2009, eichenfield2007, rosenberg2009, wiederhecker2009, wiederhecker2011}, to corral the resonators into an optical equilibrium \cite{rakich2007}, as a basis for a non-volatile optical memory \cite{varatmemory}, or to couple light with mechanical modes \cite{li2009prl, kippenberg2007, eichenfield2009}.

In most of aforementioned systems, at each operating point, the optical force is non-vanishing, i.e. it is either attractive or repulsive but never zero. The equilibrium points of structures are set by the balance between the force due to the optical fields, and the forces that are purely mechanical in origin \cite{povinelli2005, li2009, varatmemory, eichenfield2007,rakich2007, rosenberg2009, wiederhecker2009, wiederhecker2011}. It is interesting to explore whether it is possible to achieve an equilibrium just by using the optical force alone. We refer to such an equilibrium that arises purely by using optical force alone as an "optical equilibrium" for the rest of the paper. Ref. \cite{rakich2007} has shown that it is indeed possible to achieve such a pure optical equlibrium between two resonators; for a given operating frequency, the optical force has an equilibrium point; the optical force is repulsive for a small separation between the two resonators, and attractive for a large separation. Ref. \cite{rakich2007} has further demonstrated the importance of such a pure optical equilibrium for creating optical circuits that are self-tuned and self-stablized. 

In this paper, we consider the possibility of creating an optical equilibrium in the waveguide-resonator system. We analytically and numerically study the lateral force in the system of a resonator side-coupled to a feeding waveguide (example in Fig. 1a). We find that an optical equilibrium cannot occur if the resonator supports only a single resonance in the vicinity of the operating frequency. Instead, in order to have an equilibrium, the system has to be have at least two resonances that overlap in frequency. We provide detailed discussion of the requirements of these resonances in order to create an optical equlibrium in the waveguide-resonator systems.

This paper is organized as follows. In Sec. II, we analyze a waveguide-resonator system where the resonator supports a single resonance, and show analytically that no optical equilibrium can occur. We then verify this analysis with numerical simulations. In Sec. III, we proceed to two-resonance systems and explain intuitively how to achieve an optical equilibrium in these systems. We also provide simulation results of two-resonance systems which confirm our intuitive arguments. We conclude in Sec. IV.

\section{The lack of an optical equilibrium in single-resonance systems}
\subsection{Theory}
In this section, we first review the theory of the waveguide-resonator force in a single-resonance system by deriving an analytic expression for the optical force as a function of the couplings and the resonance frequency. Then, we study the existence of the optical equilibrium by studying how force changes as a function of the distance between the resonator and the waveguide.

To describe the system in Fig. 1a we consider the model as depicted in Fig. 1b. In this model, the resonator has a single resonance at the resonance frequency of $\omega_0$. The light, at a frequency $\omega$, enters the system through the waveguide, which couples to the resonator with a coupling constant $\gamma_e$. The light can exit from the resonator to the output port of the waveguide. The light can also dissipate through any loss mechanism such as material loss or radiation loss, with the loss rate $\gamma_0$. We will focus on the optical force between the input waveguide and the resonator. By using the coupled-mode theory for a travelling-wave resonance \cite{hausbook}, the field enhancement factor $\eta$ (the ratio between the resonator and input field) and the transmission coefficient $t$ (the ratio between the output and input field) are found to be
\begin{eqnarray}\label{a}
	\eta = e^{i \theta_1} \frac{ \sqrt{2\gamma_e/t_r}}{i \Delta +\gamma_0+\gamma_e}	\\ \label{t}
	t = e^{i \theta_2} \frac{i \Delta +\gamma_0-\gamma_e}{i \Delta +\gamma_0+\gamma_e}
\end{eqnarray}
where $\Delta = \omega-\omega_0$, $t_r$ is the round trip time in the resonator, and $\theta_1$ and $\theta_2$ are arbitrary phase factors depending on the positions along the waveguide where the fields are measured.

In order to find the optical force $F$ between the input waveguide and the resonator in this system, we use the theory in Ref. \cite{rakich2009},
\begin{equation}\label{rakicheqn}
	F = -\frac{1}{\omega}\sum\limits_{i} P_i \frac{\partial} {\partial d}  \phi_i 
\end{equation}
where $P_i$ is the power at each exit channel, $\phi_i$ is the output phase at that channel, and $d$ is the relevant distance. Note that the sign is opposite from Ref. \cite{rakich2009} because we use the $\exp{(+i\omega t)}$ convention. In our case, we consider the lateral force between the resonator and the waveguide, hence $d$ is the distance between the resonator and the input waveguide, with $F >0$ being a repulsive force. In our case, there are two exit channels: the output port of the waveguide, and the loss; therefore, Eq. \ref{rakicheqn} becomes
\begin{equation}\label{sumF}
	F = -\frac{P_{out}}{\omega} \frac{\partial} {\partial d}  \phi_{out}  - \frac{P_{loss}}{\omega} \frac{\partial} {\partial d}  \phi_{loss}
\end{equation}
Using Eq. \ref{a} and \ref{t} and defining the transmission $T = \left| t \right|^2$, we can write
\begin{eqnarray}
	P_{out} &=& P_{in}T \\
	\phi_{out} &=& \angle{t}\\
	P_{loss} &=& P_{in}(1-T)\\
	\phi_{loss} &=& \angle{\eta}
\end{eqnarray}
where on the last equation, we assume the loss has the same phase as the resonator field. By evaluating Eq. \ref{sumF}, we obtain
\begin{equation}\label{force}
	F = -\frac{2P_{in}}{\omega} \left(\frac{\gamma_{om}\Delta}{\Delta^2 + (\gamma_0+\gamma_e)^2}\right)  -\frac{2P_{in}}{\omega} \left(\frac{g_{om}\gamma_e}{\Delta^2 + (\gamma_0+\gamma_e)^2} \right)
\end{equation}
where
\begin{eqnarray}
	g_{om} &=& \frac{\partial} {\partial d} \omega_0\\
	\gamma_{om} &=& \frac{\partial} {\partial d} \gamma_e	
\end{eqnarray}
The results here are the same as Ref. 13 but the derivation here is different. 

At a given frequency $\omega$, a stable equilibrium occurs if there exists a resonator-waveguide separation $d_0$ such that 1) the lateral force is $0$ at $d_0$, 2) the force is attractive for $d >d_0$, and 3) the force is repulsive for $d<d_0$. Therefore, to study the possibility for a stable optical equilibrium, we need to study how the sign of force changes in $d$ at a fixed frequency $\omega$. We will show analytically that in fact, at a fixed $\omega$, the force cannot change sign as $d$ varies. As a result, neither a stable nor unstable equilibrium point can be obtained from a single resonance.

To understand how the force changes with $d$, we need to know how $\gamma_e$ and $\omega_0$ explicitly vary with $d$. Because the resonator and the waveguide couple evanescently, we expect $\gamma_e$ and $\omega_0$ to vary with $d$ as follows:
\begin{eqnarray}\label{gammae}
	\gamma_e &=& \Gamma \exp(-\kappa d) \\ \label{omega0}
	\omega_0 &=& \omega_{\infty} + \Omega \exp(-\kappa d) 
\end{eqnarray}
where $\kappa$ is the decaying constant. $\Gamma$ is a positive constant. $\omega_{\infty}$ is the resonance frequency in the absence of the waveguide. $\Omega$ is a constant that can be positive, zero, or negative, depending on the resonance's type, mode, or polarization. Eq. \ref{gammae} and \ref{omega0} will be verified numerically in the next section.

From Eq. \ref{gammae} and \ref{omega0}, we get 
\begin{eqnarray} \label{gammaom}
	\gamma_{om} &=& -\kappa \Gamma \exp(-\kappa d) = -\kappa \gamma_e\\ \label{gom}
	g_{om} &=& -\kappa \Omega \exp(-\kappa d)
\end{eqnarray}
we see that $\gamma_{om} < 0$, hence the first term of Eq. \ref{force}, which is antisymmetric with respect to the resonant frequency, shows an attractive force on the lower-frequency side and a repulsive force on the higher-frequency side. However, $g_{om}$ can be either positive or negative, so the second term of Eq. \ref{force}, which is symmetric around the resonant frequency, is repulsive if $g_{om} <0$ and attractive if $g_{om} >0$. As a result, the total force has a lineshape that is asymmetric, with the dominant sign of force depending on the sign on $g_{om}$. This is illustrated in Fig. 2 where we assume an over-coupling regime ($\gamma_e >> \gamma_0$) to achieve a large resonance contribution to the optical force \cite{ einat2011, varateit, eichenfield2007, li2009prl}. Note that the force peaks at neither $\omega_0$ nor $\omega_{\infty}$.

At each waveguide-resonator separation $d$, we now solve for the frequency $\omega_z$ where the force vanishes. Using Eqs. \ref{force} and \ref{gammae}-\ref{gom}, and after a few lines of algebra, we obtain a remarkable result:
\begin{equation}
	\omega_z = \omega_{\infty}
\end{equation}
In another word, independent of the waveguide-resonator separation, the force always vanishes at the frequency $\omega_\infty$, which is the resonance frequency of the resonator in the absence of the waveguide. Moreover, since $\omega_z$ is independent of $d$, at each frequency, the force never changes sign as a function of $d$. Therefore, one can not achieve an optical equilibrium in this system. Also, while the results here are derived for a travelling-wave single mode resonator, we note that the form of the force spectrum, i.e. Eq. \ref{force}, and the dependency of various resonance parameters on $d$, i.e. Eqs. \ref{gammae}-\ref{omega0}, apply to single-mode standing-wave resonator as well. Therefore, the main conclusion here, that one can not achieve an optical equilibrium in the waveguide-resonator lateral force using a single-mode resonator, should in general hold.

\subsection{Numerical example}

As a concrete example to support the theory in the previous section, we consider a system comprising a racetrack-shaped ring resonator side-coupled to a waveguide as shown in Fig. 1a. The waveguide and the resonator are made of silicon with permittivity $\epsilon_{Si} = 12.1$ and are surrounded by air with $\epsilon_{air} = 1$. Both the waveguide and the ring has a width of 0.2 $\mu$m. The radius of the semi-circular part of the resonator is $1.5\mu$m and the straight section is $1.1\mu$m long. The separation $d$ between the resonator and the waveguide will be varied in the range of $0.1\mu$m - $0.5\mu$m. The operating optical frequency corresponds to a free space wavelength near $1.55\mu$m. We normalize the frequency to $2\pi c/a$, where $c$ is the light speed in vacuum, and $a = 1 \mu$m. Hence a free space wavelength of $1.55 \mu$m corresponds to an angular frequency of $0.65 \times 2\pi c/a $. For simplicity, we consider a 2D system in TM mode (out-of-plane electric field and in-plane magnetic field). The simulation of this system is done using finite-element frequency-domain method with a commercial software Comsol \cite{comsol}, and the optical force is calculated by integrating Maxwell's stress tensor.

First, we verify that coupled mode theory indeed applies by plotting the transmission $T = \left| t\right|^2$ as a function of frequency $\omega$ for several $d$ in Fig. 3a. We can see that the transmission spectrum has a symmetric dip as expected. We can also deduce that $g_{om}$ is negative for this system because the resonance frequency decreases as $d$ increases. 

We then fit our parameters $\gamma_e$, $\gamma_0$ and $\omega_0$ from the transmission spectrum using Eq. \ref{t} for each value of $d$. We plot these parameters with respect to $d$ in Fig. 3b. We see that $\gamma_0$ is constant in $d$, and $\gamma_e$ and $|\omega_0-\omega_{\infty}|$ decay exponentially with approximately the same decay constant ($21.8/a$ and $21.3/a$ respectively), which verify Eq. \ref{gammae} and \ref{omega0}. We also found $\omega_{\infty} =  0.655 \times 2 \pi c/a$. 

Next, we verify the theoretical formula for the force spectrum (Eq. \ref{force}) by examining the numerically computed force spectra as shown in Fig. 3c. In this plot, the force per unit input power $F$ is normalized to the unit of $1/c$. We can see that the force spectra are indeed asymmetric, with a small attractive force on the lower-frequency side and a large repulsive force on the higher-frequency side, as expected from Eq. \ref{force} and the sign of $g_{om}$. Also, while the frequency where the maximum force occurs shifts significantly as $d$ varies, we see that the force zero $\omega_z$ does not change as $d$ changes, as expected from our theory.

To further emphasize that an optical equilibrium can not be achieved with a single resonance, we plot the sign of $F$ as a function of $\omega$ and $d$ in Fig. 3d. We can see that at large waveguide-resonator separation $d$ where the single-resonance condition is well-satisfied, the frequency $\omega_z$ where the force vanishes indeed does not change as $d$ varies, and has the value of $\omega_{\infty}$ as calculated above. As a result, no equilibrium exists. At smaller $d$, however, $\omega_z $ changes slightly as $d$ changes. This occurs because at low $d$, the linewidth of the resonance is sufficiently large such that there is additional contribution from the adjacent resonances, and as a result our assumption of having a single resonance no longer applies. Even in these cases where $\omega_z$ does vary as a function of d, this deviation in $\omega_z$ is much smaller than the linewidth at those $d$, and also smaller than the change in $\omega_0$ as $d$ changes, because the adjacent resonances are far away in frequencies. The analysis here therefore indicates that in order to achieve optical equilibrium in the waveguide-resonator system, one needs the resonator to support at least two resonances that are in close proximity to each other in frequency.

\section{Creating an optical equilibrium using two resonances}

Building upon the understanding of the force behavior of a single resonance, in this section we will present an intuitive understanding of how an optical equilibrium can be created in two-resonance system. Guided by this intuition, we then design structures that indeed achieve optical equilibrium in waveguide-resonator systems.

\subsection{Theory}

To construct a stable optical equilbrium, in Fig. 4a, consider a resonator system supporting two resonances, both with $g_{om} < 0$. We assume that the two resonances are close to each other in frequency.  For each resonance, the frequency $\omega_z$ where the force vanishes for each resonance does not vary as $d$ changes, and the force is attractive for $\omega<\omega_z$ and repulsive for $\omega>\omega_z$. Therefore, by assuming that the total force is the sum of the contributions from the two resonances, we see that the optical equilibrium can only occur at the frequencies between $\omega_z$'s of the two resonances (shaded region in Fig. 4), where the contributions from the two resonances are in opposite directions. Next, as shown in Fig. 2, a $g_{om} <0$ resonance, which has an asymmetric lineshape in its force spectrum, has a repulsive side with a larger peak amplitude, and an attractive side with a smaller peak amplitude. Also, as $d$ increases, this repulsive peak narrows significantly, while the attractive peak changes less significantly in its width. With the arrangement in Fig. 4a, the force between the two $\omega_z$'s is a sum of the contributions of the repulsive side of the lower frequency resonance, and the attractive side of the higher frequency resonance. Therefore, the way each resonance varies with $d$ as described above can lead to a change in the sign of force. The total force should change from repulsive to attractive as $d$ increases, creating a stable optical equlibrium. With a similar argument, we can see that a system with two $g_{om}>0$ resonances, as shown in Fig. 4b, results in an unstable optical equilibrium.

Guided by the intuition above, in the remaining parts of this section, we provide two concrete examples that exhibit a stable optical equilibrium.

\subsection{First example: a bumped ring resonator}
In the first example, we modify the ring resonator in Sec. II by adding a small circular bump  (radius of $0.06\mu$m) at the top of the resonator as shown in Fig. 5a-b. The ring resonator in Sec. II supports a pair of degenerate travelling-wave resonances. An incident wave in the waveguide from one side couples only to one of these two travelling wave resonances, and as a result the system can be described as a single-mode resonator. In the presence of the bump, the two travelling wave resonances in the ring couple to each other to form two standing wave resonances, with intensity patterns shown in Fig. 5a and Fig. 5b. We see that one of the standing wave resonances have an intensity node, while the other has an intensity anti-node, at the position of the bump. Both resonances now couple to the waveguide, as can be seen in the transmission spectra for a range of waveguide-resonator separation $d$ at Fig. 5c. All these spectra exhibit two resonance dips. Since both resonances shift to smaller frequencies as $d$ increases, we deduce that $g_{om} <0$ for both resonances. Therefore, the resonances in this system have the characteristics of what is required in Fig. 4a.

The force spectra for this system, for $d = 0.18\mu$m, and $d = 0.25\mu$m, were shown in Fig. 5d. We see that around the frequency $\omega_e = 0.6549 \times 2\pi c/a$ (marked with the vertical line), the system indeed exhibit a stable optical equilibrium, with the force changes from repulsive to attractive, at a fixed frequency, as $d$ increases. To emphasize this fact, we plot the force as a function of $d$ at the frequency $\omega_e$ in Fig. 5e. This clearly shows a stable optical equilbrium at $d_0 = 0.21 \mu$m; at this frequency, as $d$ moves away from $d_0$, the optical force always points towards $d = d_0$. We also plot the sign of $F$ as a function of $\omega$ and $d$ in Fig. 5f, which shows that there exists a stable optical equilibrium over the frequency range of $0.6546 \times 2\pi c/a$ to $0.6550 \times 2\pi c/a$ (shown as the bracket in the figure).

In comparing Fig. 5c and 5d, we see that in this system, at $\omega_e$, at smaller $d$, the repulsive force contribution from the lower-frequency resonance dominates over the attractive force contribution of the higher-frequency resonance. On the other hand, as $d$ increases, the contribution from the lower-frequency resonance decays faster than the higher-frequency resonance. As a result, the force changes from repulsive to attractive as $d$ increases, creating a stable optical equilibrium. In this case, the fast decay of the contribution of the lower-frequency resonance comes not only from the fact that it has $g_{om}<0$ and therefore has a fast-narrowing repulsive peak as indicated in Fig. 4a, but also that in this case, the lower-frequency resonance enters under-coupling regime (where force is small \cite{einat2011, varateit, eichenfield2007, li2009prl}) at smaller $d$ than the higher-frequency resonance. This fact can be deduced from the plot of $T$ in Fig. 5c because in an under-coupling regime, the dip in $T$ spectrum becomes shallower as $d$ increases.

\subsection{Second example: using two families of traveling-wave modes in a single ring resonator}
In this example, we use a thick ring resonator (an inner radius of $1.80 \mu$m and an outer radius of $2.15\mu$m). This resonator can support two families of travelling-wave modes, with the intensity patterns shown in Fig. 6a-b. The first-order modes (Fig. 6a) have no intensity node inside the ring. The second-order modes (Fig. 6b) have an intensity node inside the ring. By studying the force lineshape of each mode (not plotted), we found that the first-order modes have $g_{om} \approx 0$ while the second-order modes have $g_{om} < 0$. The positions of these two families of modes in frequency are shown in Fig. 6c, where the ratio of the energy inside the ring to input power ($E$), in the unit of $a/c$, is plotted as a function of frequency at $d=0.18 \mu$m. We can see that in the two frequency ranges of $0.642 \times 2 \pi c/a$ to $0.648 \times 2\pi c/a$ and $0.663 \times 2\pi c/a$ to $0.668 \times 2\pi c/a$, there are two modes, one from each of the two families, that overlap in frequency. Based on the arguments above we will therefore seek to find an optical equilibrium in these ranges. Moreover, the relative positions in frequency of the two modes are different in each range; in the first range, the second-order mode is at a higher frequency, but in the second range, the first-order mode is at a higher frequency. Therefore, these two ranges of frequency provide an interesting contrast of how different resonances affect the creation of an optical equilibrium. 

First, we consider the frequency range of $0.663 \times 2\pi c/a$ to $0.668 \times 2\pi c/a$, where the second-order mode is at lower frequency than the first-order mode as shown in Fig. 7a. In this system, the $g_{om}<0$ resonance has the lineshape that is very asymmetric with a very high repulsive peak; therefore, the contribution from this repulsive peak near $\omega_e = 0.665 \times 2\pi c/a$ decreases significantly as $d$ increases. This drop in the repulsive force, combining with the contribution from the attractive side in the lineshape of the high-frequency first-order mode, creates a stable optical equilibrium. This is shown in Fig. 7a where, near $\omega_e$, the force changes from repulsive to attractive as $d$ increases. The plot of $F$ as a function of $\omega$ and $d$ in Fig. 7b also shows a stable optical equilibrium occurring in the frequency range shown by the bracket.

Next, we consider the frequency range of $0.642 \times 2 \pi c/a$ to $0.648 \times 2\pi c/a$, where the second-order mode is at higher frequency than the first-order mode (Fig. 8). In this case, the $g_{om}<0$ resonance (second-order mode) is at higher frequency. Because the strong repulsive contribution of this resonance does not lie between the two resonances, this system lacks the large swing in the force that is required for the change in the sign of force. As a result, an optical equilibrium does not occur in this system. This is shown in Fig. 8a and 8b, where no change in the sign of force is observed.

\section{Summary}

In summary, we studied the optical equilibrium in the lateral optical force in resonator-waveguide system. We proved analytically and numerically that an optical equilibrium cannot occur in a single-resonance system. We also provide an intuitive picture, supported by numerical simulations, on the conditions for creating optical equilibrium in two-resonance systems. We believe our work will be useful in the development of self-tuned and robust optical circuits based on optomechanics. 

\section*{Acknowledgements}
This work is supported by an AFOSR-MURI program on Integrated Hybrid Nanophotonic Circuits (Grant No. FA9550-12-1-0024).

\pagebreak



\pagebreak
Figure 1
a) Schematic for a travelling-wave ring resonator. The ring and the waveguide are $0.2\mu$m wide. The radius of the semi-circular part of the resonator is $1.5\mu$m and the straight section is $1.1\mu$m long. The waveguide and the resonator have permittivity of 12.1 and are surrounded by air with permittivity of 1. The red color shows the local light intensity (proportional to the square of the electric field). b) Schematic of a general resonator-waveguide system. The resonator has a single resonance at $\omega_0$. The incident light enters the input waveguide on the port marked with a big green arrow. The coupling rate between the resonator and the input waveguide is $\gamma_e$. The resonator has an intrinsic loss rate of $\gamma_0$.

Figure 2
The force lineshapes for three regimes of $g_{om}$. The dashed lines are at a smaller waveguide-resonator distance $d$ than the solid lines. Notice that the force vanishes at the same frequency for different $d$.

Figure 3
Simulation result for the system in Fig. 1a. (The length unit $a = 1 \mu$m.) a) $T$ as a function of $\omega$ for $d = 0.2\mu$m (dash-dotted line), $d = 0.25\mu$m (dashed), and $d = 0.3\mu$m (solid). b) Fitted values of $\gamma_0$ (solid), $\gamma_e$ (dashed) and $|\omega_0-\omega_{\infty}|$ (dash-dotted) as a function of $d$. c) $F$ as a function of $\omega$ for the same $d$ as a). d) The sign of $F$ as a function of $\omega$ and $d$ (shaded means attractive, white means repulsive). The dashed line is $\omega_0$ as a function of $d$.

Figure 4
a) Force lineshape for a system with two $g_{om}<0$ resonances. The top plot is at smaller $d$ than the bottom plot. The dashed lines are the contributions from each resonance. The solid lines are the total force. The frequency range between the zeros of force is shaded. b) Same as a) but with two $g_{om}>0$ resonances.

Figure 5
a) and b) the schematic of the ring resonator in Fig. 1b with a small circular bump (radius of $0.06\mu$m). The red color shows the light intensity of the two modes. c) $T$ and d) $F$ as a function of $\omega$ for $d = 0.20\mu$m (dashed line), and $d = 0.32\mu$m (solid). e) $F$ as a function of $d$ at $\omega = \omega_e \equiv 0.655 \times 2\pi c/a$. f) The sign of $F$ as a function of $\omega$ and $d$ (shaded means attractive, white means repulsive). The bracket denotes the frequency range where an optical equilibrium occurs.

Figure 6
The schematic of the two-mode ring resonator with an inner radius of $1.80 \mu$m and an outer radius of $2.15 \mu$m. The red color shows the light intensity. a) the first-order mode. b) the second-order mode. c) The ratio of the energy inside the resonator to the input power ($E$) as a function of $\omega$ at $d = 0.18 \mu$m. Dark green arrows denote the first-order modes. Light orange arrows denote the second-order modes.

Figure 7
$F$ for a system in Fig. 7 near the frequency of $0.665c/a$. a) $F$ as a function of $\omega$ for $d = 0.06\mu$m (dashed line), and $d = 0.1\mu$m (solid). b) The sign of $F$ as a function of $\omega$ and $d$ (shaded means attractive, white means repulsive). $\omega_e = 0.655 \times 2\pi c/a$. The bracket denotes the frequency range where an optical equilibrium occurs.

Figure 8
$F$ for a system in Fig. 7 near the frequency of $0.645c/a$. a) $F$ as a function of $\omega$ for $d = 0.06\mu$m (dashed line), and $d = 0.1\mu$m (solid). b) The sign of $F$ as a function of $\omega$ and $d$ (shaded means attractive, white means repulsive).

\pagebreak
 \begin{figure}[htb] 
 \centering
 \includegraphics[width=10cm]{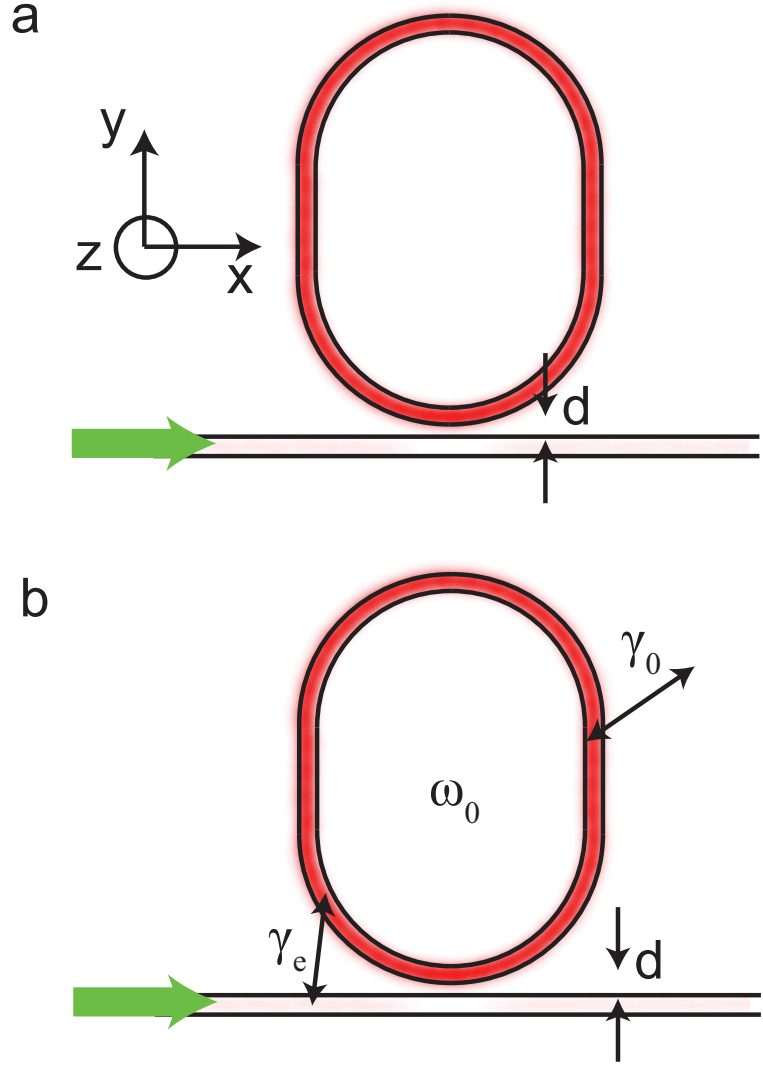}
 \caption{ a) Schematic for a travelling-wave ring resonator. The ring and the waveguide are $0.2\mu$m wide. The radius of the semi-circular part of the resonator is $1.5\mu$m and the straight section is $1.1\mu$m long. The waveguide and the resonator have permittivity of 12.1 and are surrounded by air with permittivity of 1. The red color shows the local light intensity (proportional to the square of the electric field). b) Schematic of a general resonator-waveguide system. The resonator has a single resonance at $\omega_0$. The incident light enters the input waveguide on the port marked with a big green arrow. The coupling rate between the resonator and the input waveguide is $\gamma_e$. The resonator has an intrinsic loss rate of $\gamma_0$.}
 \end{figure}

\pagebreak
 \begin{figure}[htb] 
 \centering
 \includegraphics[width=15cm]{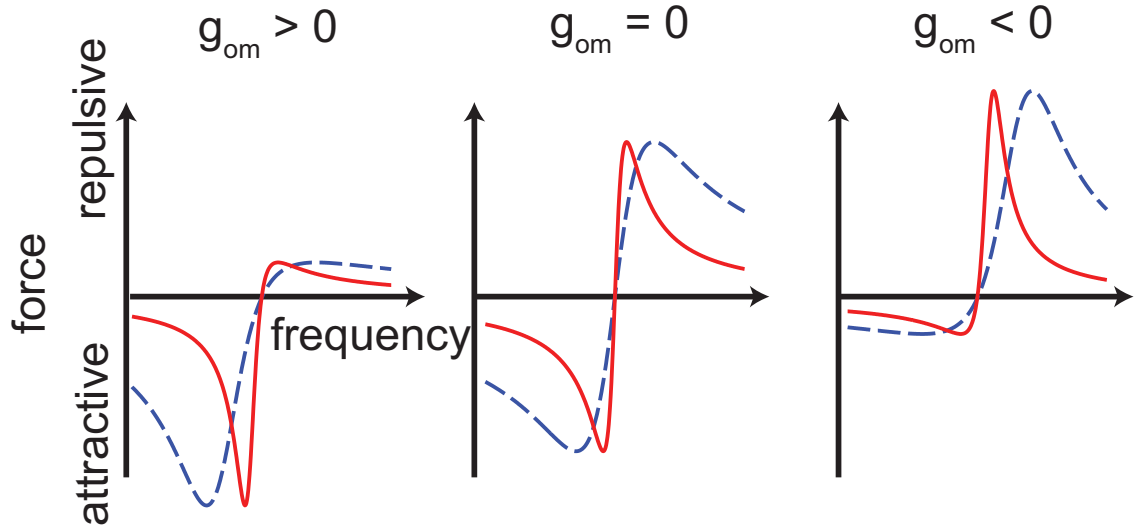}
 \caption{The force lineshapes for three regimes of $g_{om}$. The dashed lines are at a smaller waveguide-resonator distance $d$ than the solid lines. Notice that the force vanishes at the same frequency for different $d$.}
 \end{figure}

\pagebreak
 \begin{figure}[htb] 
 \centering
 \includegraphics[width=15cm]{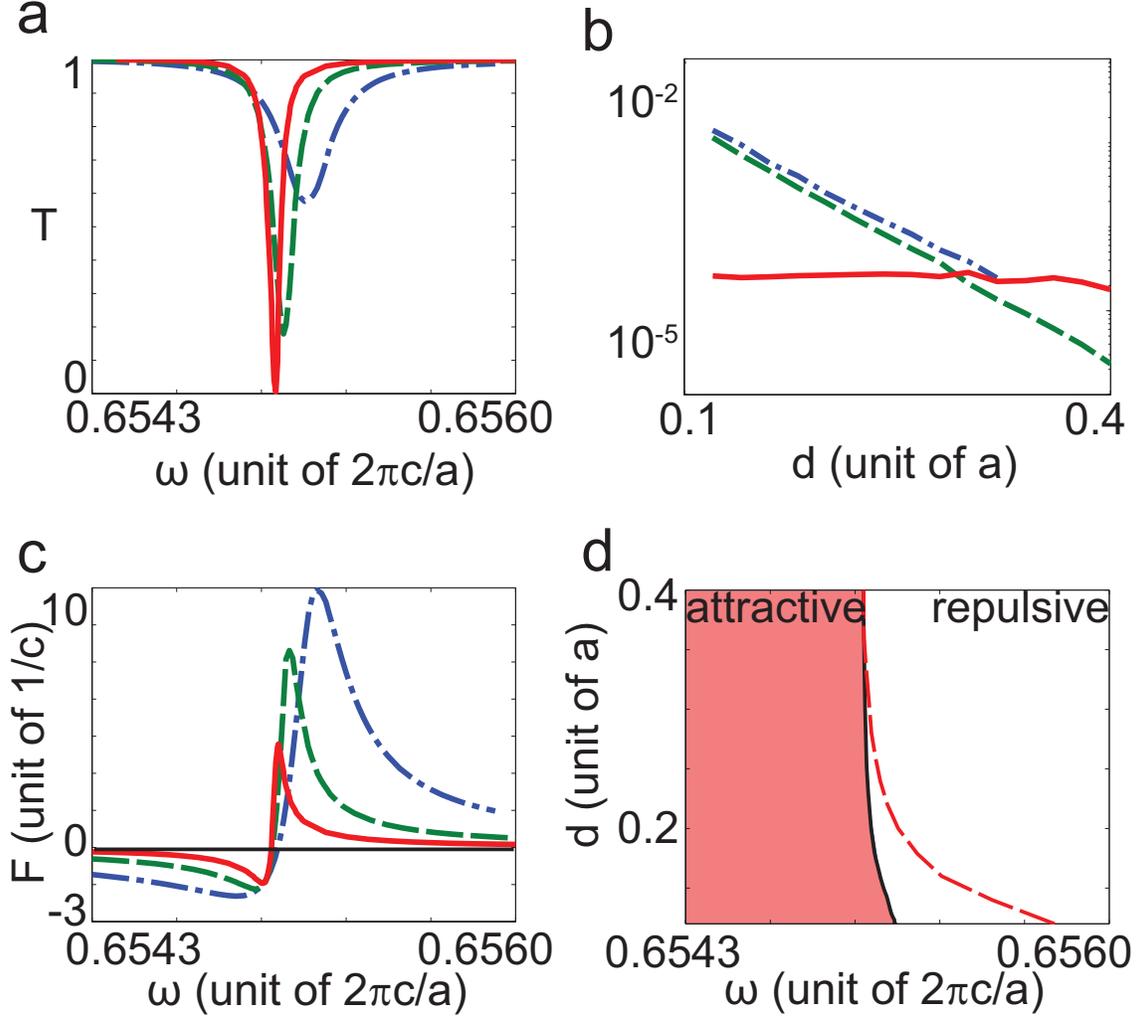}
 \caption{Simulation result for the system in Fig. 1a. (The length unit $a = 1 \mu$m.) a) $T$ as a function of $\omega$ for $d = 0.2\mu$m (dash-dotted line), $d = 0.25\mu$m (dashed), and $d = 0.3\mu$m (solid). b) Fitted values of $\gamma_0$ (solid), $\gamma_e$ (dashed) and $|\omega_0-\omega_{\infty}|$ (dash-dotted) as a function of $d$. c) $F$ as a function of $\omega$ for the same $d$ as a). d) The sign of $F$ as a function of $\omega$ and $d$ (shaded means attractive, white means repulsive). The dashed line is $\omega_0$ as a function of $d$.}
 \end{figure}

\pagebreak
 \begin{figure}
 \centering
 \includegraphics[width=15cm]{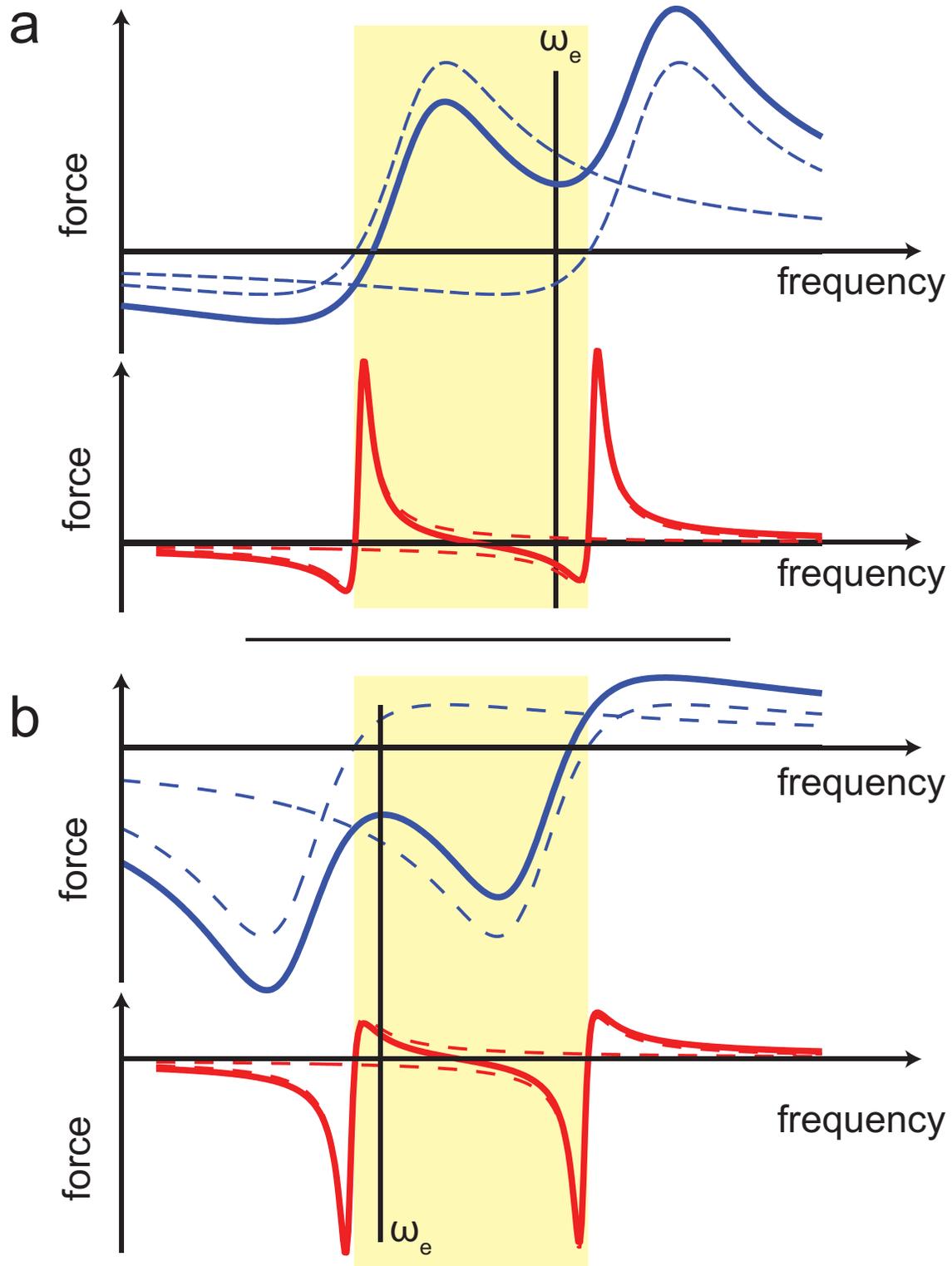}
 \caption{a) Force lineshape for a system with two $g_{om}<0$ resonances. The top plot is at smaller $d$ than the bottom plot. The dashed lines are the contributions from each resonance. The solid lines are the total force. The frequency range between the zeros of force is shaded. b) Same as a) but with two $g_{om}>0$ resonances.}
 \end{figure}

\pagebreak
 \begin{figure}[htb] 
 \centering
 \includegraphics[width=15cm]{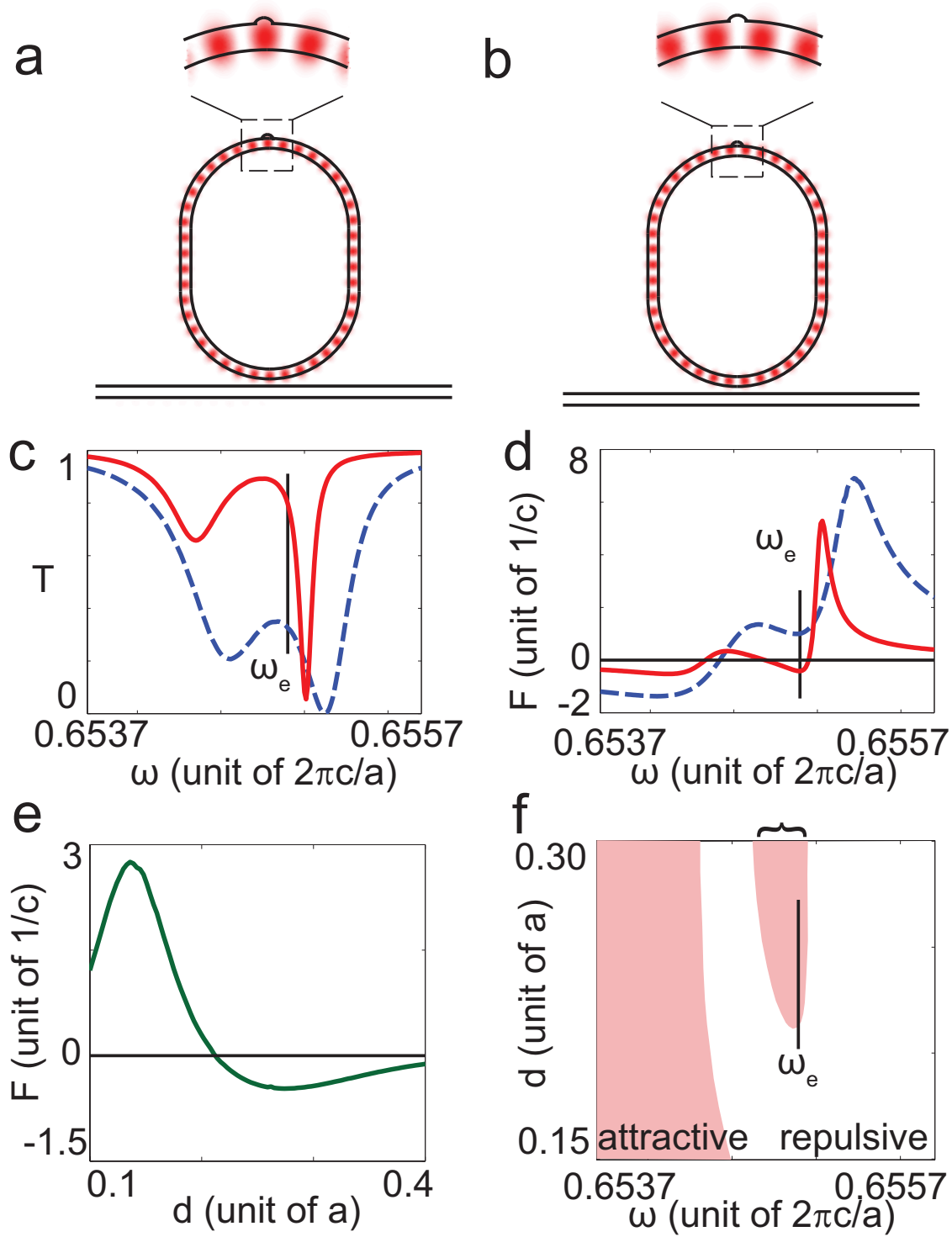}
 \caption{a) and b) the schematic of the ring resonator in Fig. 1b with a small circular bump (radius of $0.06\mu$m). The red color shows the light intensity of the two modes. c) $T$ and d) $F$ as a function of $\omega$ for $d = 0.20\mu$m (dashed line), and $d = 0.32\mu$m (solid). e) $F$ as a function of $d$ at $\omega = \omega_e \equiv 0.655 \times 2\pi c/a$. f) The sign of $F$ as a function of $\omega$ and $d$ (shaded means attractive, white means repulsive). The bracket denotes the frequency range where an optical equilibrium occurs.}
 \end{figure}

\pagebreak
 \begin{figure}[htb] 
 \centering
 \includegraphics[width=15cm]{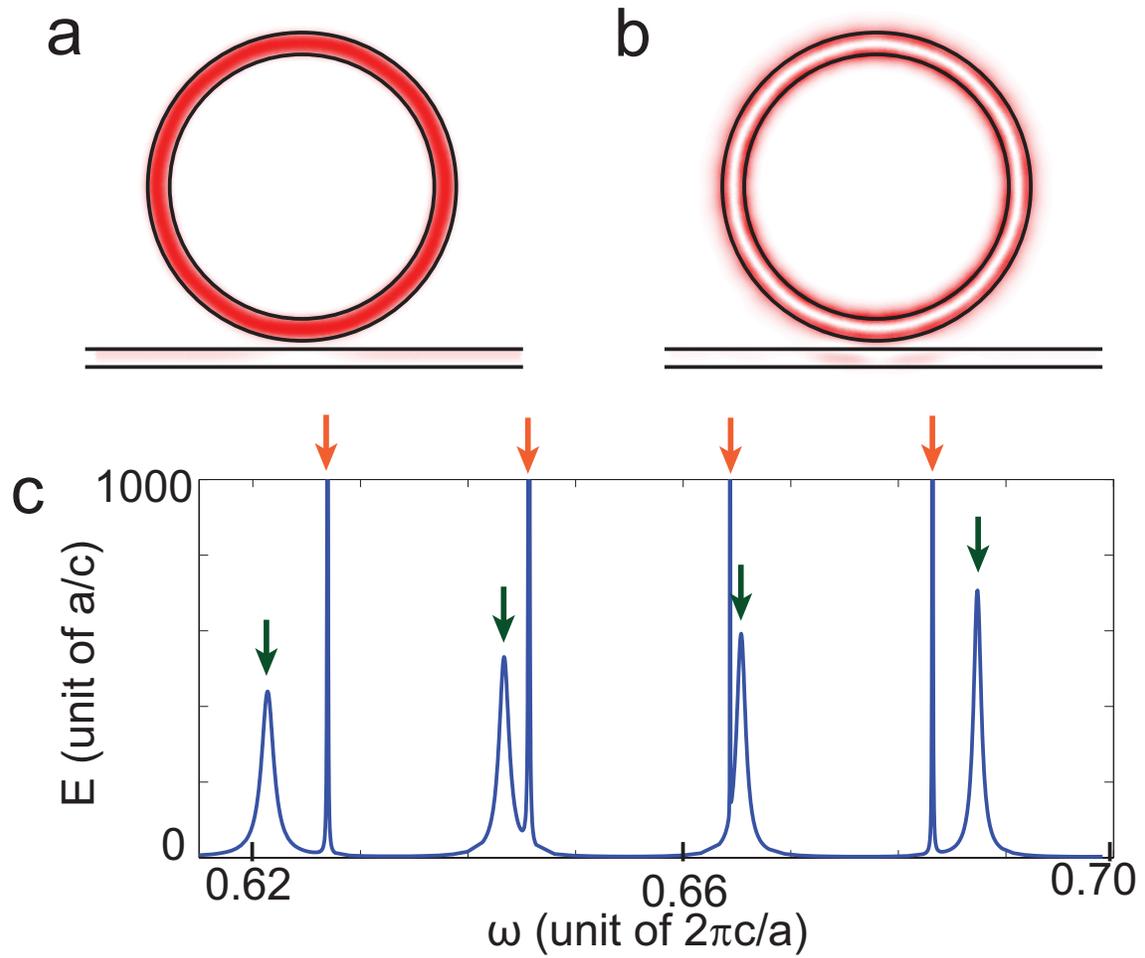}
 \caption{The schematic of the two-mode ring resonator with an inner radius of $1.80 \mu$m and an outer radius of $2.15 \mu$m. The red color shows the light intensity. a) the first-order mode. b) the second-order mode. c) The ratio of the energy inside the resonator to the input power ($E$) as a function of $\omega$ at $d = 0.18 \mu$m. Dark green arrows denote the first-order modes. Light orange arrows denote the second-order modes.}
 \end{figure}

\pagebreak
 \begin{figure}[htb] 
 \centering
 \includegraphics[width=15cm]{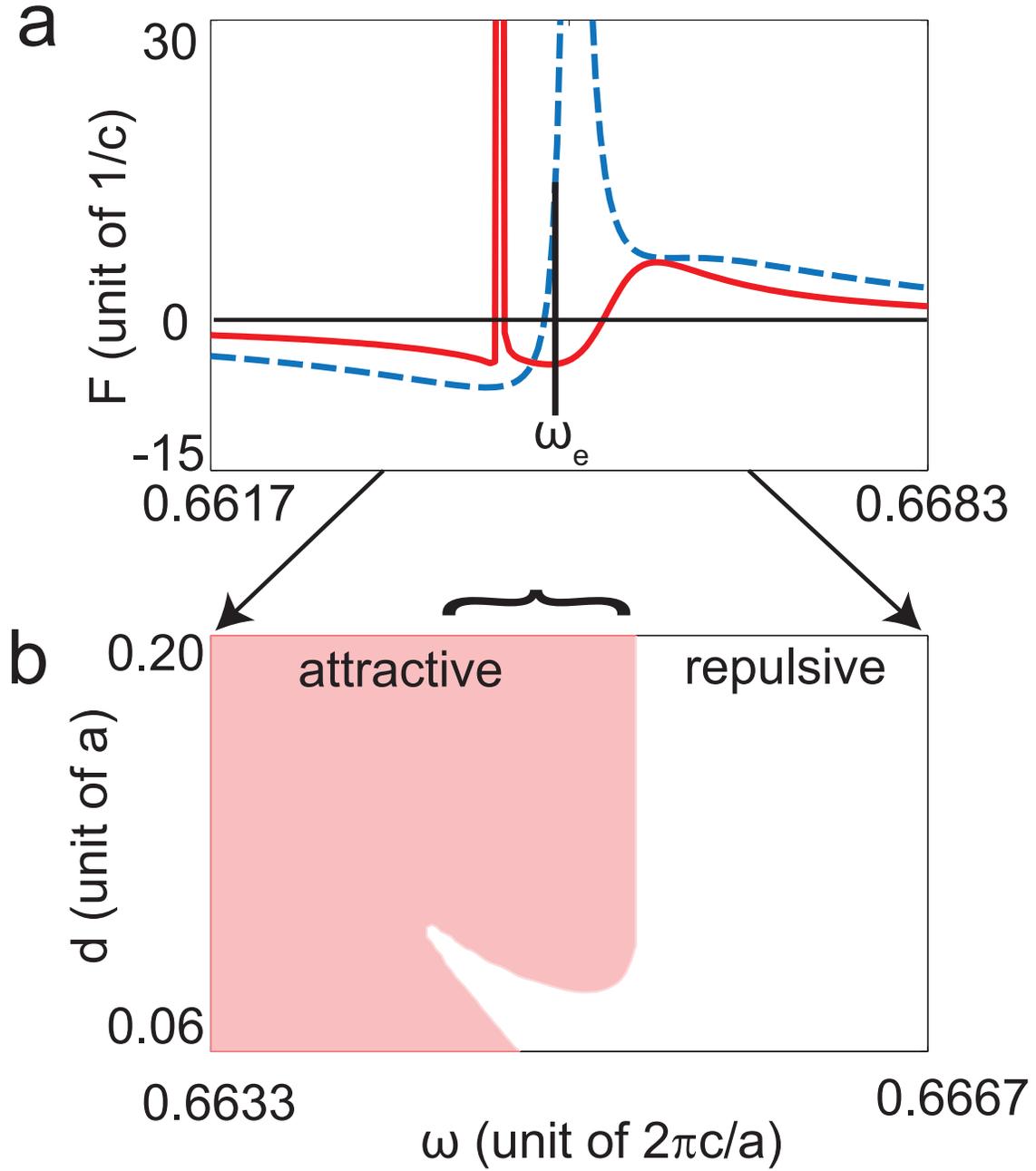}
 \caption{$F$ for a system in Fig. 7 near the frequency of $0.665c/a$. a) $F$ as a function of $\omega$ for $d = 0.06\mu$m (dashed line), and $d = 0.1\mu$m (solid). b) The sign of $F$ as a function of $\omega$ and $d$ (shaded means attractive, white means repulsive). $\omega_e = 0.655 \times 2\pi c/a$. The bracket denotes the frequency range where an optical equilibrium occurs.}
 \end{figure}
 
\pagebreak
 \begin{figure}[htb] 
 \centering
 \includegraphics[width=15cm]{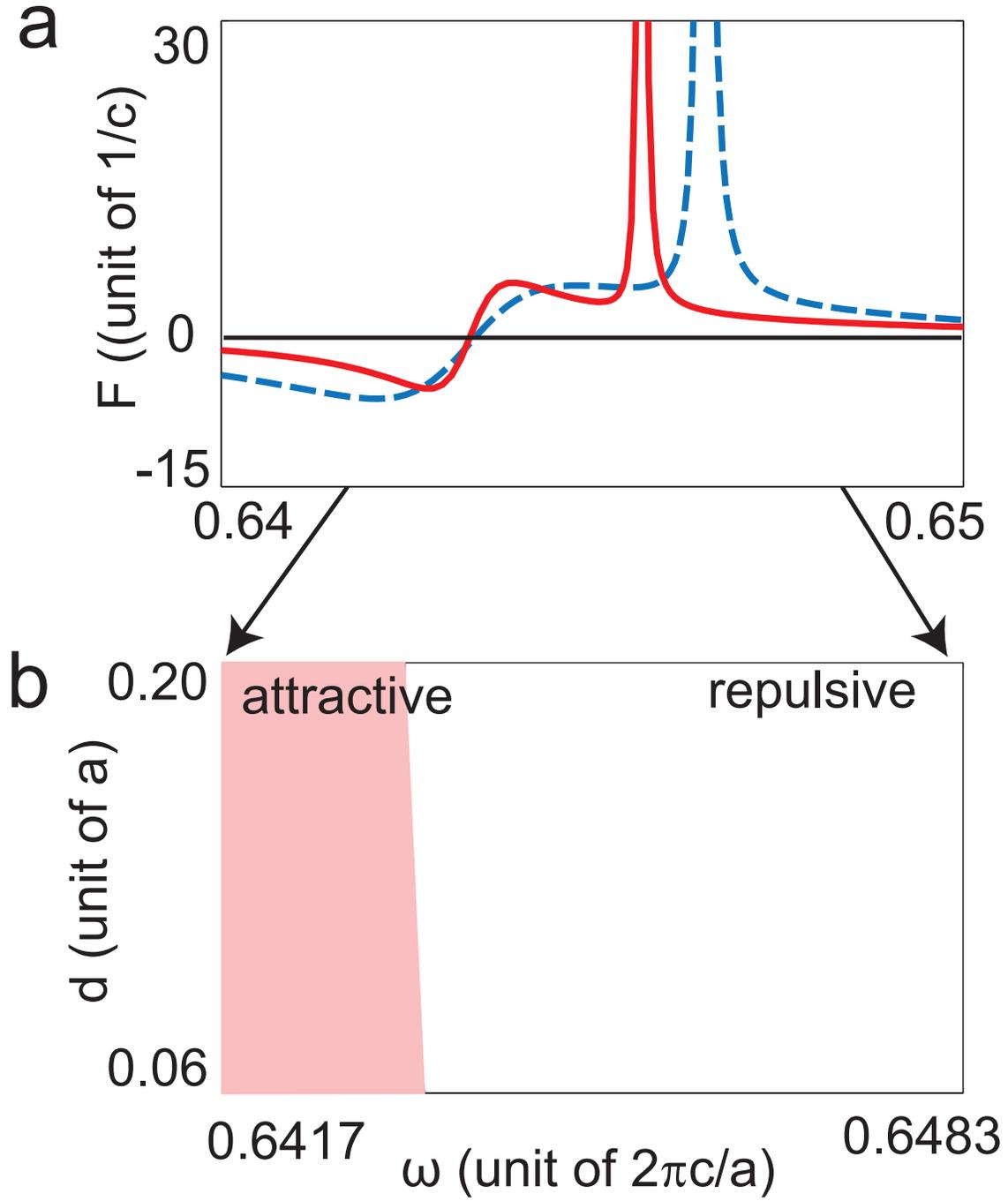}
 \caption{$F$ for a system in Fig. 7 near the frequency of $0.645c/a$. a) $F$ as a function of $\omega$ for $d = 0.06\mu$m (dashed line), and $d = 0.1\mu$m (solid). b) The sign of $F$ as a function of $\omega$ and $d$ (shaded means attractive, white means repulsive).}
 \end{figure}

\end{document}